\documentclass[english,aps,twocolumn,amssymb,prl,showpacs]{revtex4}
\usepackage[T1]{fontenc}
\usepackage[latin9]{inputenc}
\usepackage{amsmath}
\usepackage{graphicx}
\usepackage{amssymb}
\usepackage{esint}

\makeatletter
\@ifundefined{textcolor}{}
{%
 \definecolor{BLACK}{gray}{0}
 \definecolor{WHITE}{gray}{1}
 \definecolor{RED}{rgb}{1,0,0}
 \definecolor{GREEN}{rgb}{0,1,0}
 \definecolor{BLUE}{rgb}{0,0,1}
 \definecolor{CYAN}{cmyk}{1,0,0,0}
 \definecolor{MAGENTA}{cmyk}{0,1,0,0}
 \definecolor{YELLOW}{cmyk}{0,0,1,0}
 }

\usepackage{epsfig}
\@ifundefined{showcaptionsetup}{}{%
\PassOptionsToPackage{caption=false}{subfig}}
\usepackage{subfig}
\makeatother

\usepackage{babel}

\begin{document}

\title{Photoinduced helical metal and magnetization in two-dimensional electron systems with spin-orbit coupling}

\author{Teemu Ojanen$^{1,2}$}
\email[Correspondence to ]{teemuo@boojum.hut.fi}
\author{Takuya Kitagawa$^2$}
\affiliation{$^1$Low Temperature Laboratory, Aalto University, P.~O.~Box 15100,
FI-00076 AALTO, Finland }
\affiliation{$^2$Physics Department, Harvard University, Cambridge, Massachusetts 02138, USA}

\date{\today}
\begin{abstract}
Helical metals realized at the surfaces of topological insulators
have recently attracted wide attention due to their potential
applications in spintronics.
In this paper we propose to realize helical metals
through the application of THz light on
common two-dimensional semiconductors and discuss their observable properties.
We show that the application of circularly polarized light enables coherent manipulation of magnetization.
Moreover, for a range of chemical potentials the system behaves as a helical metal, exhibiting a large anomalous Hall conductivity and associated magnetoelectric effect. Proposed dynamical engineering of material properties through light in much-studied
materials opens new perspectives for future applications.
\end{abstract}
\maketitle
\bigskip{}

\emph{Introduction}--
Physical properties of materials can be tailored by the application of
electromagnetic fields \cite{fausti, miyano, syzranov}.
Recently, the engineering of topological band structures through
application of light has been proposed
in various semiconductors \cite{lindner1, lindner2, dora}, graphene \cite{kitagawa1, kitagawa2, gu} and
cold atom systems \cite{jiang}.
These proposals suggest that physical behavior of matter
is not solely determined by static microscopic properties
but can be strongly affected by an external dynamical control.
In this paper we demonstrate how to realize a helical metallic state from a common two-dimensional semiconductor structure with a Rashba or Dresselhaus
spin-orbit coupling
through application of light.

The existence of helical surface states is
one of the most striking consequence of recently discovered three
dimensional topological insulators (TI) \cite{kane2}.
The surface of TI forms a novel $\mathbb{Z}_2$ metallic phase which
evades localization even in the presence of strong disorder, as long
as the time-reversal symmetry is not broken. In addition, spin and momentum of the charge carriers are perfectly correlated. Such unique
features of helical metals are of great interest from the point of view of future electronics and spintronics applications. Recently it was proposed that similar helical properties could be realized in a two-dimensional electron gas (2DEG) with a
strong Rashba spin-orbit coupling \cite{sau1}. In this proposal, 2DEG
is placed in the proximity of a ferromagnetic material, which opens a Zeeman gap in the center of the Brillouin zone. Provided that the Fermi level can be tuned to lie in the gap, the system behaves as a helical metal. A realization of such helical metal in well-studied semiconductor materials is highly desirable for technological developments. However, fabrication of 2DEGs with strong spin-orbit coupling and a sizable Zeeman gap is experimentally challenging and has not been achieved yet.

Here we propose an alternative method to realize helical metals by application of circularly polarized THz radiation in 2D semiconductor structures with a strong spin-orbit coupling. Typical materials that could be employed for this purpose are Indium-based compounds such as InAs, GaInAs/GaAlAs structures, II-VI semiconductor compounds or surface structures with strong Rashba coupling. Circularly polarized light breaks the time-reversal symmetry in the material, acting as an effective Zeeman field thus circumventing the difficulty of fabricating 2DEG structures in the proximity of magnetic materials. We provide a detailed account on three striking manifestations of the photoinduced state. First of all, the state exhibits a coherent optomagnetic effect resulting in a light polarization-dependent out-of-plane magnetization as response to the irradiation. Secondly, when the Fermi energy lies in the photoinduced gap, the state displays anomalous Hall conductivity of the order of $e^2/2h$. The anomalous Hall effect is accompanied by a magnetoelectric effect, an inplane magnetization parallel to the applied dc electric field. These phenomena can be experimentally probed by the Faraday-Kerr effect and transport measurements, providing a characterization of the photoinduced helical metal. A helical metal in the proximity of superconductivity is a crucial ingredient of topological superconductors \cite{sau1, lutchyn, oreg}, so our work paves the way for photoinduced topological superconductivity in proximity coupled systems.

\begin{figure*}[t]
\centering

 \begin{tabular}{c  c  c  c} 
\label{a} \includegraphics[width=0.5\columnwidth, clip=true]{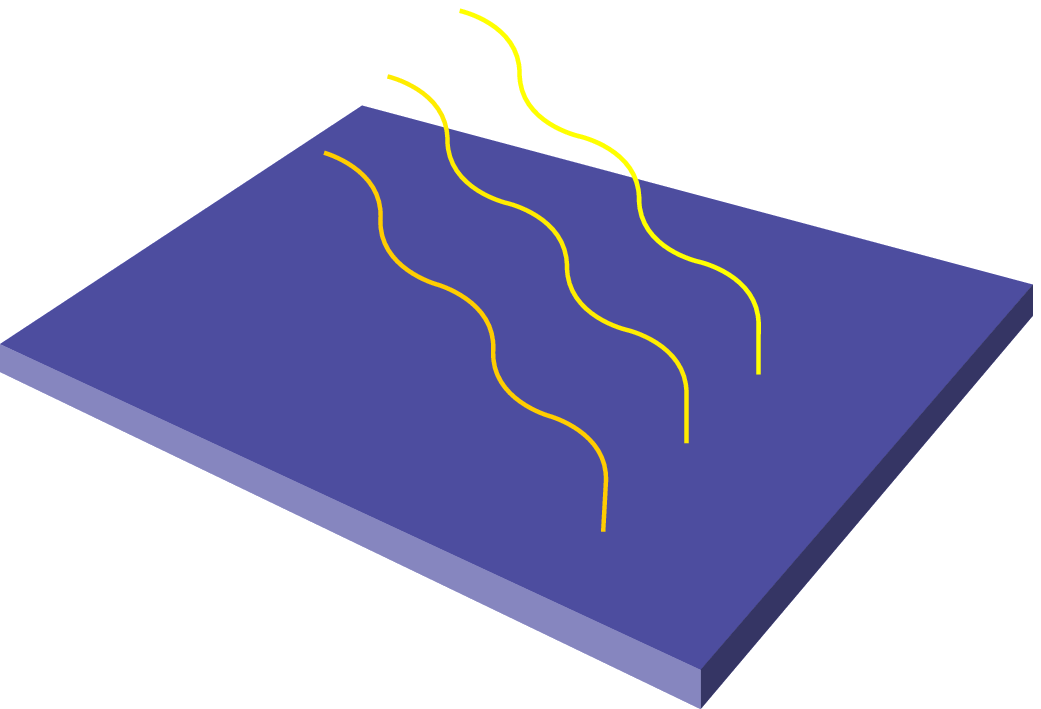}
 &\label{b} \includegraphics[height=0.5\columnwidth]{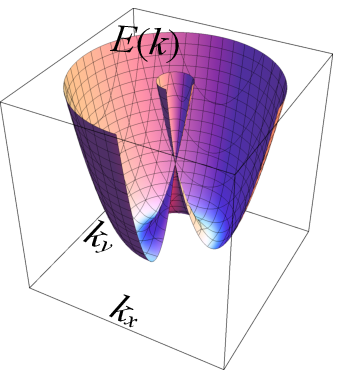}
 &\label{c} \includegraphics[height=0.5\columnwidth]{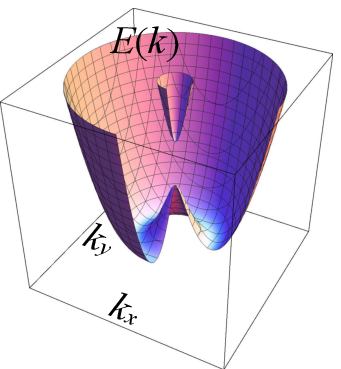} &
\label{d} \includegraphics[height=0.45\columnwidth]{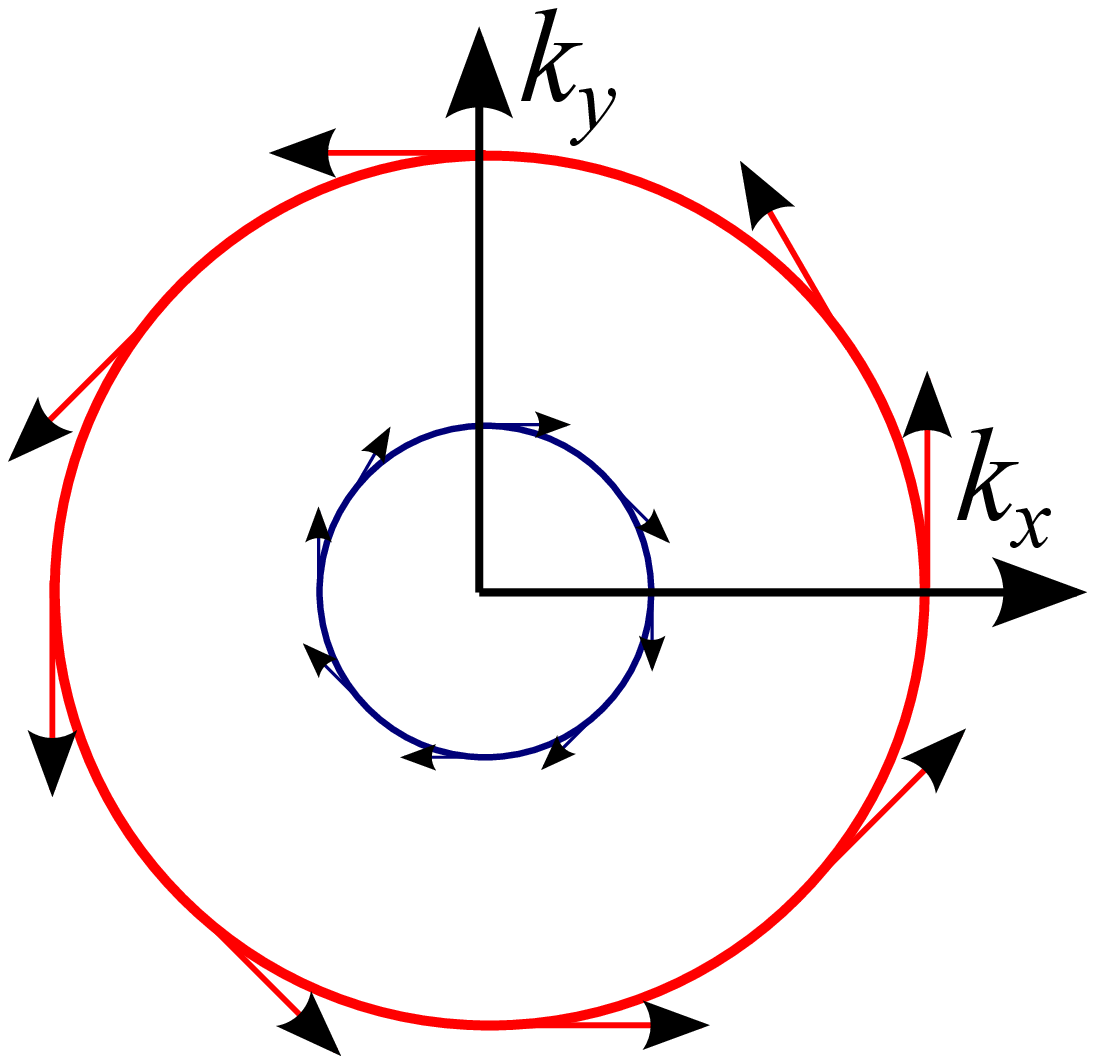}
 \\ 
(a) & (b) & (c) & (d) \\
 \end{tabular}
\caption{(a): System under study, a spin-orbit coupled 2DEG under circularly polarized THz irradiation (b): Spectrum of the 2DEG with a Rashba coupling (c): Spectrum of the effective Hamiltonian in the presence of irradiation. The degeneracy at k=0 is lifted, leading to a finite gap between the bands. (d): Fermi surface of the spectrum (b) for a generic Fermi energy. Arrows indicate spin directions, being opposite for the two bands and perpendicular to momenta. In the presence of irradiation the inner Fermi surface vanishes for Fermi energies in the gap and the carriers have a unique spin-momentum relation. Spins also acquire out-of-plane components proportional to the gap. }   \label{g}
\end{figure*}
\emph{Model}-- We are studying a 2DEG system under external irradiation as depicted in Fig.~\ref{g} (a). Our starting point is a free 2DEG with a Rashba spin-orbit coupling described by the Hamiltonian
\begin{align}\label{hzero}
H_0=\frac{\hbar^2k^2}{2m}+\alpha(k_x\sigma_y-k_y\sigma_x),
\end{align}
where $k_{x,y}$ label the momentum components in the sample plane and $\sigma_{x,y}$ are the usual Pauli matrices. Below we analyze a pure Rashba coupling, although the results remain qualitatively the same in the presence of a linear Dresselhaus term $\beta(k_x\sigma_x-k_y\sigma_y)$  as long as $\alpha\neq\beta$. The spectrum is illustrated in Fig.~\ref{g} (b) where it can be seen that there exists two Fermi surfaces for a generic chemical potential.
In the presence of circularly polarized irradiation the Hamiltonian is modified by the minimal substitution $k\to k-\frac{eA}{\hbar}$
where the field is included through the time-dependent vector potential $A=\frac{E_0}{\Omega}\left(\mathrm{cos}(\Omega\, t), \mathrm{sin}(\Omega\,t) \right)$.
Our primary interest lies in the time-averaged dynamics of the system, so it is convenient to define a time-averaged propagator
 $G^r(\omega)=\frac{1}{T}\int_0^T dt'\int_{-\infty}^{\infty} dt e^{i\omega t}G^r(t+t',t').$
The retarded propagator in the absence of time-dependent field is given by the standard expression $G_0^r(\omega)=\frac{1}{\omega-H_0+i\eta}$. In the frequency regime $\frac{\hbar eE_0k_F}{m\Omega}$, $\frac{\alpha eE_0}{\hbar\Omega}\ll \hbar\Omega$ where the field-dependent terms are treated perturbatively. The self-energy  arising from the external field has two contributions
\begin{align}\label{self}
\Sigma^r(\omega)=\sum_{\pm}\frac{H_{\mp1}H_{\pm1}}{\omega-H_0\mp\Omega+i\eta}\approx\frac{[H_1,H_{-1}]}{\hbar\Omega},
\end{align}
where $H_{\pm1}=\frac{1}{T}\int_0^Tdt H e^{\pm i\Omega t}$ \cite{kitagawa1}. The last form is valid for the states close to the Fermi surface $\omega-H_0\ll\Omega$ . The propagator becomes $G^r(\omega)=\frac{1}{\omega-H_{\rm{eff}}+i\eta}$, where
\begin{align}\label{heff1}
H_{\rm{eff}}=H_0+\frac{\left[H_{1},H_{-1}\right]}{\hbar\,\Omega}=H_0+\Delta\sigma_z
\end{align}
with $\Delta=\frac{(\alpha eE_0)^2}{(\hbar\Omega)^3}$. Thus, the field-dressed propagation can be described by an effective Hamiltonian containing a band renormalization from the real part of the self-energy. These self-energy corrections describe virtual two-photon emission-absorption processes. The role of real absorption, potentially important in the presence of disorder and phonons, is discussed below. The effective Hamiltonian (\ref{heff1}) has eigenvalues $E^{\pm}(k)=\frac{\hbar^2k^2}{2m}\pm\sqrt{\alpha^2k^2+\Delta^2}$ and eigenstates
\begin{align}\label{eig}
|\psi^\pm\rangle=\frac{e^{i\boldsymbol{k}\cdot\boldsymbol{r}}}{\sqrt{2}}\left(\begin{array}{c}
                                        \mp ie^{-i\phi_k}\frac{\rm{sin}\,\theta_k}{\sqrt{1\mp\rm{cos}\,\theta_k}} \\
                                         \sqrt{1\mp\rm{cos}\,\theta_k}\\
                                      \end{array}
                                    \right),
\end{align}
where $\phi_k=\mathrm{arctan}\frac{k_y}{k_x}$, $\mathrm{sin}\,\theta_k=\frac{\alpha k}{\sqrt{\alpha^2k^2+\Delta^2}}$, and $\mathrm{cos}\,\theta_k=\frac{\Delta}{\sqrt{\alpha^2k^2+\Delta^2}}$. The essential influence of the field is to introduce a gap $2\Delta$ between the bands at $k=0$ as depicted in Fig.~\ref{g} (c). The gap is crucial for the existence of a helical state and leads to various striking properties as discussed below.
When $\Delta\ll \alpha k$ the states (\ref{eig}) resemble to the left and right-handed eigenstates of a massless Dirac equation describing the states in TI surfaces and graphene. When the Fermi energy lies in the gap, only the lowest band is populated at low temperatures and the system behaves effectively as a helical metal. Further insight of the states can be gained by evaluating the expectation values of the spin operators:
$\langle S_x \rangle_k^{\pm}= \langle \psi^{\pm}|\,S_x |\psi^{\pm}\rangle=\mp\frac{\hbar}{2} {\rm sin} \theta_k\, {\rm sin}\, \phi_k $,
$\langle S_y \rangle_k^{\pm}=\pm\frac{\hbar}{2} {\rm sin} \theta_k\, {\rm cos}\, \phi_k$ and
$\langle S_z \rangle_k^{\pm}=\pm\frac{\hbar}{2} {\rm cos}\, \theta_k$. These results confirm that the inplane component of spin, as illustrated in Fig.~\ref{g} (d), is always perpendicular to the momentum and opposite for the two bands. A finite gap $\Delta$ also results in an out-of-plane spin component which has an opposite sign for the two bands. Below we analyze physical properties and experimental signatures of the photoinduced state.

\emph{ Photoinduced out-of-plane magnetization}--
As shown above, the optically induced gap $\Delta$ leads to a coherent out-of-plane spin component that has different sign for the two bands. Here we show that this leads to a net out-of-plane spin polarization that can be tuned by the external field. The magnetization per unit area for a band is
\begin{align}\label{}
 \rho_ {S_z}^{\pm}=\frac{\langle S_z\rangle^{\pm}}{A}=\frac{1}{A}\sum_k \langle S_z \rangle_k^{\pm}n_k^{\pm}
\end{align}
where $n_k^{\pm}$ is the Fermi distribution of the band and $A$ is the sample area. Since the out-of plane spin component for the upper band $|\psi^+\rangle$ and the lower band $|\psi^-\rangle$ have different signs, the magnetization arising from states located outside the band gap is expected to mostly cancel out. The density of states and the spin directions of the two bands combine in such a way that the cancelation is in fact exact. Thus the net magnetization $\rho_ {S_z}=\rho_ {S_z}^{-}+\rho_ {S_z}^{+}$ arises \emph{completely} from the lower band states lying in the gap and  can be expressed as
\begin{align}\label{mag}
\rho_ {S_z}=\frac{1}{A}\sum_{k\in C} \langle S_z \rangle_k^{-}n_k^{-}
\end{align}
where $C$ denotes the set of momenta for which $E^-(k)$ lie in the gap. 
For large enough chemical potentials $\mu > |\Delta|$ that both bands are occupied, the magnetization at zero temperature is given by
\begin{align}\label{magnet}
 \rho_ {S_z}=
 -\frac{\hbar}{2}\left(\frac{\Delta m}{\hbar^2\pi}\right),
\end{align}
which is, as explained above, independent of the chemical potential. Temperature corrections to (\ref{magnet}) are exponentially small $\mathcal{O}(e^{-\beta(\mu-|\Delta|)})$ at low temperatures. The gap $\Delta$ changes sign when the circular polarization is reversed, a property that can be used to distinguish the photoinduced contribution from possible other sources of magnetization independent of the polarization of light. Though not a direct evidence of the helical spin structure, the out-of-plane magnetization provides a simple proof of the photoinduced band renormalization and can be observed through a Faraday-Kerr effect \cite{kato} as discussed below. Importantly, observation of the optomagnetic effect \emph{does not} require gating the Fermi energy in the gap or temperatures small compared to $\Delta$.

\emph{ Anomalous Hall conductivity and magnetoelectric effect}--  One of the hallmarks of a helical TI surface metal is a half-integer quantized Hall conductance when time-reversal symmetry is broken by a Zeeman field $m$ perpendicular to the surface. The resulting Hall conductance $\sigma_H=\frac{m}{|m|}\frac{e^2}{2h}$ per surface cannot be realized in any strictly 2D insulator where the Hall conductance is always an integer multiple of $e^2/h$. Below we show that the photoinduced state, where the time-reversal is broken by a finite gap $\Delta$, exhibits an anomalous Hall effect. When the chemical potential is tuned to lie in the gap, the Hall conductivity takes values of the order of the half-quantized value $\frac{e^2}{2h}$. The Hall effect is accompanied by a magnetoelectric effect, an inplane magnetization as a response to the inplane dc electric field.

Starting from the Kubo formula and a two-band Hamiltonian such as (\ref{heff1}), one can derive a convenient expression for the Hall conductivity
\begin{equation}\label{kubo}
\sigma_{xy}=\frac{e^2}{\hbar}\frac{1}{8\pi}\int d^2k \epsilon^{lmn}\hat{d}_l\partial_{k_x}\hat{d}_m\partial_{k_y}\hat{d}_n\left(n^+_k-n^-_k\right),
\end{equation}
where $\boldsymbol{\hat{d}}=\boldsymbol{d}/|\boldsymbol{d}|$, $\boldsymbol{d}=(-\alpha k_y,\alpha k_x, \Delta)$ and $\epsilon^{lmn}$ is the antisymmetric tensor \cite{qi}. Evaluation of the formula at zero temperature yields
\begin{equation}\label{hall1}
\sigma_{xy}=\frac{e^2}{2h} \left(\frac{\Delta}{\sqrt{\alpha^2 k^2_{F+}+\Delta^2}}-\frac{\Delta}{\sqrt{\alpha^2 k^2_{F-}+\Delta^2}} \right),
\end{equation}
where $k_{F\pm}$ denote the Fermi momenta of the two bands. In the helical case when the upper band is unoccupied the Hall conductivity becomes
\begin{equation}\label{hall2}
\sigma_{xy}=\frac{e^2}{2h} \left(\frac{\Delta}{|\Delta|}-\frac{\Delta}{\sqrt{\alpha^2 k^2_{F-}+\Delta^2}} \right).
\end{equation}
This result reveals the remarkable similarity of the half-quantized Hall effect of a TI surface and the anomalous Hall effect of the artificial 2d helical metal. The first term in the brackets corresponds to the half-quantized Hall conductivity, the sign of which is determined by the sign of $\Delta$. The appearance of the second nonuniversal correction term can be traced to the fact that the system is gapless so the lower band crosses the Fermi surface and cuts off the integral (\ref{kubo}) at the Fermi momentum. The value of the Hall conductivity approaches $\frac{e^2}{2h}$ when $\Delta/\alpha k_{F-}$ approaches zero. The Hall conductivity (\ref{hall2}) follows from a nonvanishing Berry curvature of the lower band and indicates that the time reversal symmetry is broken.  Since spin and momentum are coupled, the Hall current gives rise to an inplane spin accumulation. The dc electric field driving the Hall current is perpendicular to the velocity of the carriers, which in turn is perpendicular to their spin, implying that the inplane magnetization associated to the current is \emph{parallel} to the electric field. The spin density can be calculated by evaluating a linear response formula similar to (\ref{kubo}) resulting in
\begin{equation}\label{inpl}
\rho_ {S_y}=\frac{\hbar^2 e}{2\alpha}\sigma_{xy}E_y,
\end{equation}
where $E_y$ is the bias dc electric field. The magnetization $(\ref{inpl})$ as a response to the electric field provides a strong evidence of the helical nature of the system. The sign of the Hall conductivity and the associated spin accumulation can be inverted by reversing the polarization of the irradiation, a feature that can be utilized to distinguish these phenomena from other sources of Hall current that are independent of the polarization of irradiation.


\emph{Physical realization}-- In the derivation of the effective Hamiltonian (\ref{heff1}) we considered a clean noninteracting system driven by a time-dependent field. In real systems there are always impurities and interactions present imposing restrictions to the parameters. In the presence of impurities the propagator acquires a finite imaginary part given by the inverse lifetime $G_0^r(\omega)=\frac{1}{\omega-H_0+i\frac{\tau}{2}^{-1}}$. The form of Eq.~(\ref{self}) suggests that the band renormalization remains unaffected as long as $\Omega\gg \tau^{-1}/2$. This statement can be substantiated by considering self-energy diagrams contributing to $G^r(\omega)$ in the presence of disorder. Impurity vertices interrupting a high-energy propagator between $H_1$ and $H_{-1}$ ($\propto 1/\Omega$) give rise to $1/\Omega\tau$ factors which strongly suppress the contribution. As a consequence, the leading contribution come from diagrams where there is no impurity vertices on high-energy propagators. Thus, in the regime $\Omega\gg \tau^{-1}/2$ the time-averaged dynamics can be approximated by that of particles described by $H_{\mathrm{eff}}$ in a random potential. Inelastic interactions also restrict the lifetime but at low temperatures $T\sim 1$ K close to the Fermi surface the impurity scattering poses the most stringent condition for the driving frequency. Typically in GaAs and related InGaAs and InAs 2DEGs  $\tau\sim 0.1- 1\times 10^{-12}$s \cite{arnone, kabir, song} so the frequency should be $\Omega/2\pi\gtrsim 1-10$ THz to observe the field-induced band renormalization.

Large spin-orbit couplings have been reported in indium-based III-V semiconductors making them promising candidates for physical realization. For example, in InGaAs/InAlAs 2DEG systems the coupling constant can be as large as $\alpha=0.3$ eV\AA \, \cite{choi}. The observed value of the effective mass $m=0.05m_e$ in this structure implies that one can achieve a photoinduced gap  $2\Delta/k_B=10$ K by intensities corresponding to fields $E_0=2-60$ kV/cm (when $\Omega=1-10$ THz). In order to realize the helical metal, chemical potential must be tuned to lie in the gap. That indicates $E_F=2\,m\alpha^2/\hbar^2=1.2\,\mathrm{meV}$ (placing the Fermi level in the middle of the gap). These values suggest that a helical metal could be observed in this material at temperatures of few Kelvin. For the above parameters the optically induced out-of-plane magnetization (\ref{magnet}) is $\rho_ {S_z}\sim 10^{14}\, \mathrm{spins/m^2}$ which is large enough to be detected by a Kerr rotation \cite{kato}. Further, the anomalous Hall conductivity (\ref{hall2}) becomes $\sigma_{xy}=0.7\times\frac{e^2}{2h}$ and the inplane magnetization from the associated magnetoelectric effect $(\ref{inpl})$ is $\rho_ {S_y}\sim 10^{14}\, \mathrm{spins/m^2}$ for an inplane electric field $E_y=0.1$ kV/cm. Instead of employing III-V semiconductors, one could also consider II-VI or surface structures \cite{ast} where the Rashba coupling can be an order of magnitude larger. In these systems the gap would be two orders of magnitude larger for the same intensity, thus enabling exploration of the helical regime at significantly higher temperatures provided that the driving frequency exceeds the inverse lifetime of the carriers. Various optically pumped THz lasers operating in the relevant frequency regime reach (and for some frequencies exceed) 100 mW power which is sufficient for generating 1-100 $\mu \mathrm{m}^2$ area of helical 2DEG.

As noted above, the band renormalization leading to the effective Hamiltonian (\ref{heff1}) arises from virtual two-photon processes \cite{kitagawa1}. In the presence of impurities and phonons the electron system acquires a finite real part of conductivity enabling a real absorption. Absorption events excite single electrons after which they relax in a complicated cascade of an optical phonon emission and, eventually, by electron-electron collisions and an acoustic phonon emission. This process will heat up the electron system which generally equilibrates to a higher temperature than the lattice. The experiments performed in THz regime with comparable intensities at low temperatures show that the absorption and heating of the 2DEG is significantly reduced compared to the estimates deduced from the low-frequency mobility \cite{arnone, kabir, song}. Also experiments demonstrate that at cryostat temperatures of the order of 1K the effective electron temperature is elevated only by a tiny fraction of Kelvin while coherence properties are preserved \cite{arnone}. In the light of these results the method we propose to realize helical liquid seems realistic.

\emph{Conclusion}--
In this work we introduced a possibility to realize a coherent photoinduced magnetization and a helical 2DEG by employing circularly polarized THz radiation in systems with a spin-orbit coupling. This method is applicable to a variety of different materials and avoids completely the need for magnetic fields and magnetic materials. We showed that the helical phase exhibits a large anomalous Hall conductivity and associated magnetoelectric effect which results in an inplane magnetization parallel to applied dc electric fields. The photoinduced state can be realized and its consequences observed by employing established optical and transport techniques.

The authors would like thank Eugene Demler and Jay D. Sau for valuable discussions. One of the authors (T.O.) would like to thank Academy of Finland for support.

\end{document}